\documentclass[prd,12pt,preprintnumbers,nofootinbib,showpacs]{revtex4}
\usepackage{epsfig,amsmath,amssymb}
\def\KeyWord#1{$\backslash$\IfColor{$\!\!$\textRed{#1}\textBlack}{#1}$\!\!$}
\usepackage{epsfig,amsmath}

\def\Q1{{\bf Q}_1}

\begin{document}
\preprint{\bf FIRENZE-DFF - 427/07/2005}

\title{A NJL-based study of the QCD critical line}
\author{A. Barducci, R. Casalbuoni, G. Pettini, L. Ravagli}

\affiliation{Department of Physics, University of Florence and
INFN
Sezione di Firenze \\
Via G. Sansone 1, I-50109, Sesto F.no, Firenze, Italy}
\email{barducci@fi.infn.it, casalbuoni@fi.infn.it, pettini@fi.infn.it, ravagli@fi.infn.it}

\begin{abstract}                

We employ a 3 flavor NJL model to stress some general remarks about the QCD critical line.
The dependence of the critical curve on $\mu_q=(\mu_u+\mu_d)/2$ and $\mu_I=(\mu_u-\mu_d)/2$ is discussed. The quark masses are varied to confirm that, in agreement with universality arguments, the order of transition depends on the number of active flavors $N_f$. The slope of the critical curve vs. chemical potential is studied as a function of $N_f$. We compare our results with those recently obtained in lattice simulations to establish a comparison among different models.

\end{abstract}
\pacs{ 11.10.Wx, 12.38.-t,  25.75.Nq}  \maketitle

\section{INTRODUCTION}
\label{int}

In recent years, the study of the QCD phase diagram by means of numerical lattice simulations has improved considerably. 
In particular, the presence of a critical ending point, first discovered within microscopical effective models \cite{Barducci:1989wi,Halasz:1998qr,Rajagopal:2000wf}, appears to be a solid feature \cite{Fodor:2001pe,Fodor:2004nz}, althought its exact location along the critical curve is still controversial. Moreover, the incoming realization of LHC enhances the interest of the scientific community in this kind of problem.\\
The main problem regarding the lattice study of QCD phase diagram is related to the so called sign problem; the fermionic determinant is not positive definite at finite baryon chemical potential, and therefore, to avoid this unwelcome feature, some suitable tricks are needed. The most commonly used between them are the study of QCD at imaginary chemical potential \cite{D'Elia:2002gd,deForcrand:2002ci,deForcrand:2003hx}, the reweighting procedure \cite{Fodor:2001pe} and Taylor expansion in $\mu/T$ \cite{Allton:2002zi,Ejiri:2003dc}.
A possible way of avoiding the sign problem is to consider QCD at finite isospin chemical potential $\mu_I$: in this case, the fermion determinant is real and positive definite, and standard Monte Carlo simulations are allowed \cite{Kogut:2002zg,Gupta:2002kp}. Futhermore, the regime of finite $\mu_I$ can be also studied within a class of effective models, and the results can be compared with those obtained on the lattice in order to check the consistency of different approaches.
Two properties are remarkable and are of interest at $\mu_I\neq 0$: pion condensation and the splitting of critical curves related with light flavors. The former has been the subject of several studies within different models, starting from effective Lagrangians \cite{Son:2000xc,Loewe:2002tw}, random matrices \cite{Klein:2003fy}, NJL \cite{Toublan:2003tt,Barducci:2004tt} and ladder-QCD \cite{Barducci:2003un} and so apparently a model-independent phenomenon. However, pion (and kaon \cite{Barducci:2004nc}) condensation would not be accessible from Heavy Ion experiments, but could regard the physics of compact stars.
The latter question is slightly more controversial: the role played by instantons seems to be crucial to determine whether values of $\mu_I$ obtainable in experiments can produce the separation of critical lines \cite{Frank:2003ve}.\\
The aim of this paper is to offer an overview of results concerning the behaviour of the critical line, obtained in the NJL model and directly comparable with recent lattice analyses. In its strong simplicity, the NJL model recovers the basic structure of non perturbative dynamics ruling the problem.
Therefore, it can be trusted as a good toy model for the study of QCD phase diagram.

\section{The model}
\label{sec:physics}

Let us now consider the Lagrangian of the NJL model with three
flavors $u,d,s$, with current masses $m_u=m_d\equiv m$ and $m_s$
and chemical potentials $\mu_u,\mu_d,\mu_s$ respectively

\begin{eqnarray}
{\cal {L}}&=& {\cal {L}}_{0}+{\cal {L}}_{m}+{\cal {L}}_{\mu}+{\cal {L}}_{4}+{\cal {L}}_{6}\nonumber\\
&=&{\bar{\Psi}}i{\hat{\partial}}\Psi-{\bar{\Psi}}~\underline{\mathcal{M}}~\Psi~+~
\Psi^{\dagger}~\underline{\mathcal{A}}~\Psi~+~{G}\sum_{a=0}^{8}\left[\left(
{\bar{\Psi}}\lambda_{a}\Psi
\right)^{2}+\left({\bar{\Psi}}i\gamma_{5}\lambda_{a}\Psi\right)^{2}
\right]\\
&+&K\left[\mbox{det}{\bar{\Psi}(1+\gamma_5)\Psi+\mbox{det}\bar{\Psi}(1-\gamma_5)\Psi}\right]\nonumber \label{eq:njlagr}
\end{eqnarray}
where
\begin{equation} \Psi=\left(
\begin{array}{c} u\\ d\\s
\end{array} \right),~~~~~\underline{\mathcal{A}}=\left(
\begin{array}{c} \mu_u\\0\\0
\end{array} \begin{array}{c} 0\\ \mu_d\\0
\end{array}\begin{array}{c} 0 \\ 0\\ \mu_s
\end{array}\right),
~~~~\underline{\mathcal{M}}=\left(
\begin{array}{c} m\\0\\0
\end{array} \begin{array}{c} 0\\ m\\0
\end{array}\begin{array}{c} 0 \\ 0\\ m_s
\end{array}\right)
 \end{equation}

\bigskip\noindent $\underline{\mathcal{M}}$ is the current quark mass matrix which is
taken diagonal and $\underline{\mathcal{A}}$ is the matrix of the quark chemical
potentials. As usual $\lambda_0=\sqrt{\displaystyle{{2\over3}}}~
\mbox{\bf{I}}$ and
$\lambda_{a}~$, $~a=1,...,8~$ are the Gell-Mann matrices.

The 't Hooft determinant term, for the three flavors case, corresponds to a six fermion interaction. By working at the mean field level, the six fermion term can be recast into an effective four-fermion one. In such a way the Lagrangian (\ref{eq:njlagr}) reduces to the usual NJL Lagrangian, apart from a redefinition of the four-fermion coupling constant G into a new set of effective ones, taking into account the flavor mixing arising from the 't Hooft term \cite{Hatsuda:1994pi,Klevansky:1992qe}

Therefore, because we are dealing with four fermion interactions only, we can calculate the effective potential by using
the standard technique to introduce bosonic (collective) variables
through the Hubbard-Stratonovich transformation and by integrating
out the fermion fields in the generating functional.\\
If we limit ourselves to consider the three scalar condensates, and the pseudoscalar condensate in the light quark sector only, the one-loop effective potential we get is:
\begin{equation}
\label{eq:poteff} V=\frac{\Lambda^2}{8G}
(\chi_u^2+\chi_d^2+\chi_s^2)-\Lambda^3\frac{K}{16G^3}~\chi_u~\chi_d~\chi_s+\frac{\Lambda^2}{4G}(1-\frac{K\Lambda\chi_s}{8G^2})(\rho_{ud}^2)+V_{\mbox{log}}
\end{equation}
\vskip0.5cm
where

\begin{eqnarray} \label{eq:vlog}
V_{\mbox{log}}=&& -{1\over\beta}\sum_{n=-\infty}
^{n=+\infty}\int{d^{3}p\over (2\pi)^{3}} ~\mbox{Tr log} \left(
\begin{array}{ccc}
h_u & -\gamma_5 ~\Lambda~\rho_{ud}~(1-\frac{K\Lambda\chi_s}{8G^2})~&0\\
\gamma_5 ~\Lambda~\rho_{ud}~(1-\frac{K\Lambda\chi_s}{8G^2}) & h_d & 0\\
0& 0 & h_s
\end{array}
\right)
\nonumber\\
\\
&&~~~~~~~~\ \ \ \ \ \ \ \ \ \ \ \ ~h_f=(i\omega_n+\mu_f)\gamma_0~-~\vec{p}\cdot\vec{\gamma}~-~
M_f\nonumber
\end{eqnarray}
In eq. (\ref{eq:vlog}) $\mbox{Tr}$ means trace over Dirac, flavor and color indices
and $\omega_{n}=(2n+1)\pi/\beta$ are the Matsubara frequencies.
The dimensionless fields $\chi_{f}$ and $\rho_{ud}$ are
connected to the scalar and pseudoscalar condensates respectively by the following relations

\begin{eqnarray} \label{eq:fields}
\chi_f &=& - ~4G~{\langle{\bar{\Psi}}_f\Psi_f\rangle\over \Lambda}\nonumber\\
\\
\rho_{ud} &=& -~2G
~{\langle{\bar{u}}\gamma_{5}d-{\bar{d}}\gamma_{5}u\rangle\over
\Lambda}\nonumber
\end{eqnarray}
and are variationally determined at the absolute minimum of the
effective potential. The constituent quark masses are
\begin{equation}
M_i=m_i+\Lambda \chi_i-\Lambda^2\frac{K}{G^2}\frac{\chi_j~\chi_k}{8} ~~~(i\neq j \neq k)
\end{equation}

Since the model is non renormalizable, we have to introduce the hard cut-off $\Lambda$ on the three-momentum.

Through this letter, when we consider the physical case (with realistic values of meson masses and decay consant), we assume for the parameters the same values as in ref. \cite{Hatsuda:1994pi}
\begin{equation}\label{param1}
\Lambda=631.4~\mbox{MeV};~~~G~\Lambda^2=3.67;~~~K~\Lambda^5=-9.29;\\
\end{equation}
\begin{equation}\label{param2}
 ~~~~\hat{m}\equiv\frac{m_u+m_d}{2}=5.5~\mbox{MeV};~~m_s=135.7~\mbox{MeV}~~
\end{equation}

 To investigate regimes different than the real one, with a varying number of $N_f$ massless flavors, we will treat quark masses as free parameters, by keeping coupling and cut-off scale fixed as in eq. (\ref{param1}). 

In the following, it will turn out to be more convenient to introduce the following linear combinations of chemical potentials:

\begin{equation}
\mu_q=(\mu_u+\mu_d)/2;~~~\mu_I=(\mu_u-\mu_d)/2;
\end{equation}

The quark chemical potential $\mu_q$ is just one third of the baryon chemical potential $\mu_q=\mu_B/3$.
Following the analyses previously performed within the chiral Lagrangian
approach \cite{Son:2000xc,Loewe:2002tw}, and NJL \cite{Barducci:2004tt,He:2005nk} and ladder-QCD \cite{Barducci:2003un} studies,
we expect a superfluid phase with condensed pions when
the isospin chemical potential $\mu_I$ exceeds a critical value $\mu_I^C$ ($\mu_I^C=m_{\pi}/2$ at $\mu_q=T=0$).

\section{Behaviour of critical lines}

\subsection{Critical temperature dependence on baryon/isospin chemical potentials}
\begin{center}
\begin{figure}[htbp]
\includegraphics[width=14cm]{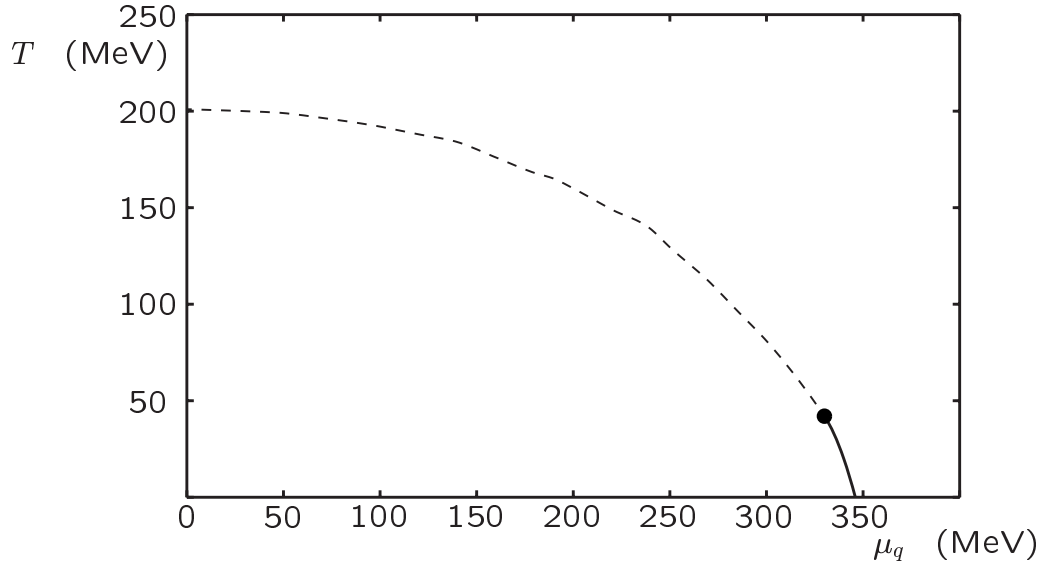}
\caption{\it Phase diagram in the plane ($\mu_q,T$), for the physical 2+1 case; $\mu_I$ and $\mu_s$ are set to zero. Dashed/solid lines indicate cross-over/first order transitions; consequently, the dot in the picture labels the critical ending point.} \label{fig:DiagTmu2+1}
\end{figure}
\end{center}

\begin{center}
\begin{figure}[htbp]
\includegraphics[width=14cm]{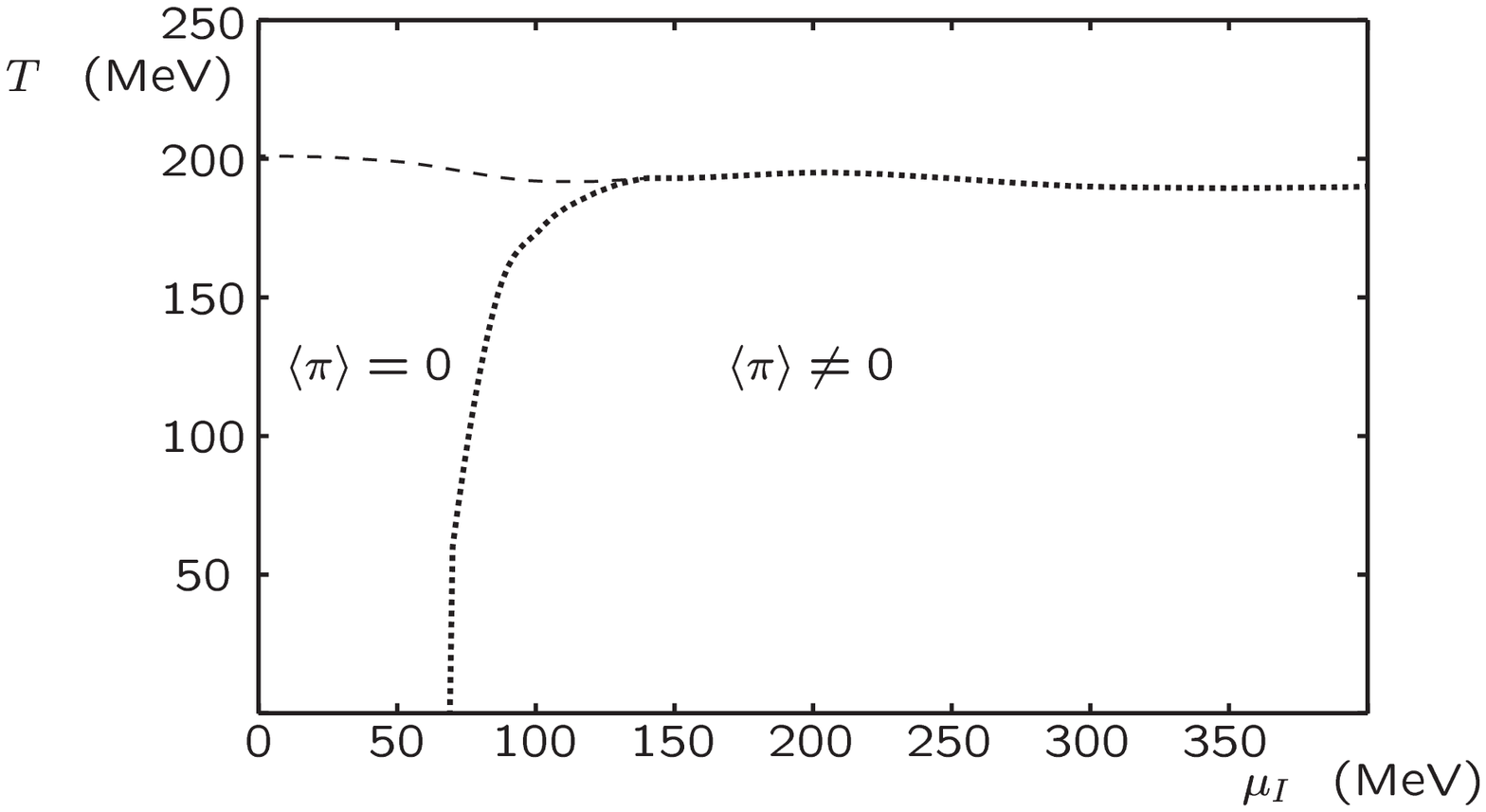}
\caption{\it Phase diagram in the plane $(\mu_I,T)$, for the physical 2+1 case; $\mu_q$ and $\mu_s$ are set to zero. The dashed line indicates a cross-over transition for the scalar condensates, whereas dotted line stands for genuine second order transition for the pion condensate. In the regions of a non-vanishing pion condensate, the discontinuous behaviour of the scalar condensate turns into a continuous one, therefore the critical ending point is not present in this case.} \label{fig:DiagTmui}
\end{figure}
\end{center}

The aim of this section is to shed some light on the physics of QCD at finite baryon chemical potential ($\mu_q=\mu_B/3$), by comparing the physics at $\mu_q\neq 0$ and $\mu_I=0$ with that at $\mu_q=0$ and $\mu_I\neq 0$. In fact, the latter case can be studied on the lattice by means of standard importance sampling techniques. The connection of these two regimes could give a deeper understanding of the sign problem in the fermion determinant, and provide us with some procedure to check present simulations and possibly improve numerical algorithms.\\
Rigorous QCD inequalities at non zero chemical potential have been proposed to try to resolve this question in \cite{Cohen:2004qp}; the enigma why, at $T=0$, there exists a critical value for chemical potentials below which the system lies in its ground state has been called "silver blaze problem". A possible solution of this problem, by using $1/N_c$ expansion, has been proposed in \cite{Cohen:2004mw,Toublan:2005rq}. A recent study of phase quenched QCD (a theory where the absolute value of the fermion determinant is taken) has been performed in \cite{Splittorff:2005wc}.\\
We will consider here the dependence of the critical temperatures $T_c(\mu_q)\equiv T_c(\mu_q; \mu_I=0)$ and $T_c(\mu_I)\equiv T_c(\mu_I; \mu_q=0)$, for low chemical potentials, obtained by a mean-field analysis of the NJL model.
Obviously, mesons and baryons (and di-quarks) carry different spin and charges, and their properties depend differently on $\mu_q$ and $\mu_I$; for these reasons, one could expect the two curves $T_c(\mu_q)$ and $T_c(\mu_I)$, when starting by the same value at $\mu_q=\mu_I=0$, to be different, at least in the regime where bound states heavily influence the thermodynamics of the system. On the other hand, when the free-energy is mainly ruled by the constituent quarks, there is no reason to expect a dependence of the critical curve on the signs of chemical potentials (which will fix the sign of the total charges associated with the system). \\
Here at the mean-field level, the effect of bound states is considered when we admit the formation of a pion condensate. Actually, in agreement with chiral models analyses, in the NJL model the pion effective mass dependence on chemical potentials can be analitically computed \cite{He:2005nk}; the result is that the charged pions chemical potential is exactly the double of the isospin chemical potential. For this reason, as $\mu_I$ is higher than some critical value ($m_{\pi}/2$ at $T=0$), a pion condensate starts to form; a similar effect happens when $\mu_q$ is higher than the critical value for di-quark condensation, which is expected to occur at values $\mu_q>300\div400~\mbox{MeV}$ (of course, before di-quark condensation, for $\mu_q\sim~m_N/3$ and low temperatures there should be the liquid-gas transition for the nucleons). Since in this paper we are interested mainly in the regime of relatively small chemical potentials (lower than $\sim 200~\mbox{MeV}$) we will neglect the latter possibility.\\
In fact, when the pion condensate is zero, the mean field effective potential is symmetric under $\mu_u\rightarrow-\mu_u, ~\mu_d\rightarrow-\mu_d$; this implies that $\mu_q\leftrightarrow\pm\mu_I$ is a symmetry of the problem. Therefore, for zero $\rho$, the two curves $T_c(\mu_q)$ and $T_c(\mu_I)$ have the same analytical dependence.\\
In Fig. \ref{fig:DiagTmu2+1}, \ref{fig:DiagTmui} the phase diagrams in ($\mu_q,T$), ($\mu_I,T$) spaces are shown; starting from the common value $T_0=201~\mbox{MeV}$, corresponding to $\mu_q=\mu_I=0$, the cross-over curves coincide up to the value of about $150~\mbox{MeV}$ for both chemical potentials. For higher $\mu_I$ (and temperatures lower than $\sim 200~\mbox{MeV}$) we are in the condensed pions phase; in agreement with \cite{He:2005nk} this regime will persist until $\mu_I\sim 860~\mbox{MeV}$, before the saturation regime takes place. On the other hand, if we follow the cross-over line $T_c(\mu_q)$, we will find, as expected, a critical ending point for $\mu_q=330~\mbox{MeV},T=42~\mbox{MeV}$, and a line of first order transitions for higher $\mu_q$.\\
Summarizing, at this level of calculation (mean field), until isospin chemical potential is lower than the critical value for pion condensation the curves $T_c(\mu_q)$ and $T_c(\mu_I)$ have the same analytical expression $T_c(\mu_q)=T_c(\mu_I)$. This could be interesting in the attempt to extend lattice results from $\mu_I\neq0$ and $\mu_q=0$ to $\mu_q\neq 0$ and $\mu_I=0$ (at least in the region of low chemical potentials).
Our conclusions appear to be in agreement with the authors of ref. \cite{Toublan:2004ks} (both from the lattice and from a hadron renonances gas model) .

\subsection{Order of the transition by varying $m_s$}

\begin{center}
\begin{figure}[htbp]
\includegraphics[width=14cm]{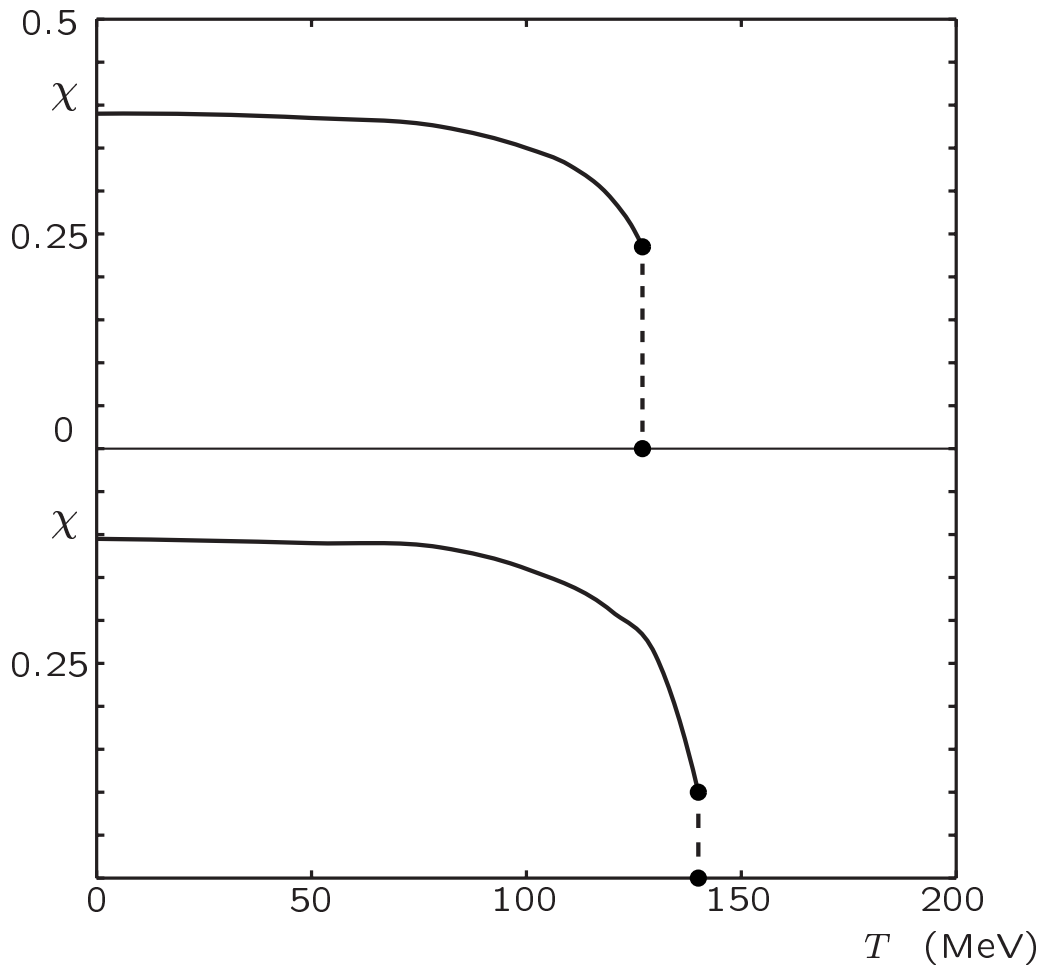}
\caption{\it Behaviour of the light quark scalar condensates as a function of temperature at zero chemical potentials and $m_u=m_d=0$, for $m_s=0$ (upper picture) and ${ m_s=8~\mbox{MeV}}$  (lower picture). In the upper picture, $\chi\equiv\chi_u=\chi_d=\chi_s$, in the lower $\chi\equiv\chi_u=\chi_d$. When the strange quark mass exceeds the critical value $m_s^C=10~\mbox{MeV}$, the discontinuous behaviour turns into a genuine phase transition.} \label{fig:CondTms8}
\end{figure}
\end{center}

\begin{center}
\begin{figure}[htbp]
\includegraphics[width=14cm]{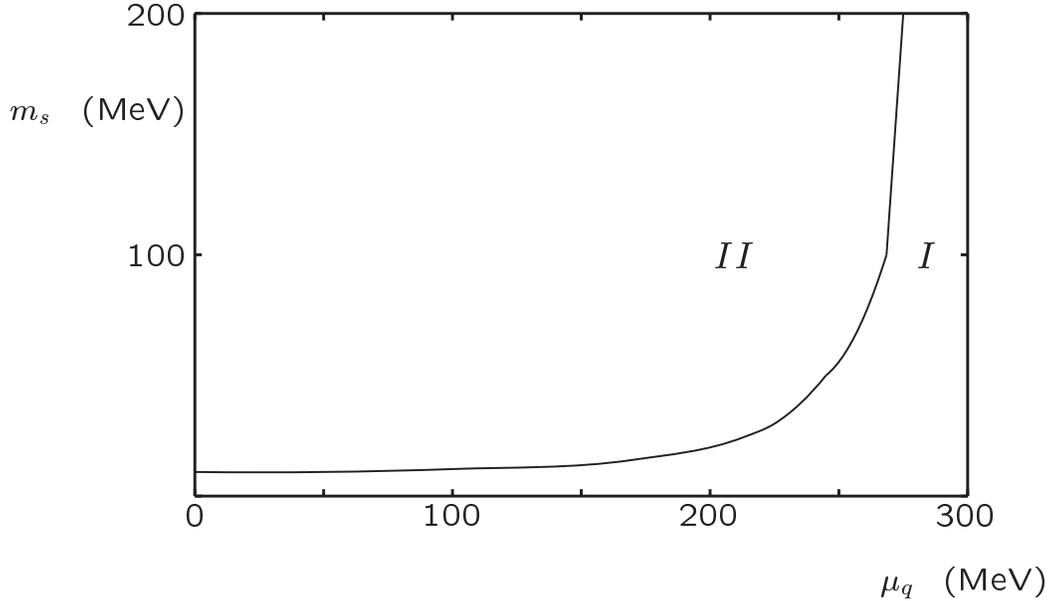}
\caption{\it Phase diagram in the ($\mu_q, m_s$) space, at $m_u=m_d=0$ and $\mu_I=\mu_s=0$. For every region in the diagram, the label $I$/$II$ means whether, by increasing temperature and starting from $T=0$, the transition is of first or second order. Actually, the line in the diagram separating the two different regions follows the critical point by varying $m_s$.} \label{fig:Diagmusmu}
\end{figure}
\end{center}

It is generally assumed from universality arguments \cite{Pisarski:1983ms} and lattice analyses, \cite{Laermann:2003cv}, that the order of the phase transition by increasing temperature at zero density can change as the 
quark mass values are varied. If in the realistic case (with physical quark masses) the zero density transition is expected to be a cross-over, lattice analyses seem to show a first order transition when the three light quark masses are small enough. In particular, by taking $m_u=m_d=0$, there should be a critical value for $m_s$ below which the transition turns into a discontinuous one: different lattice approaches find $m_s^C$ to be half of the physical value of the strange quark mass \cite{Brown:1990ev} or $m_s^C\sim 5\div10 ~ m_{u,d}~(\mbox{physical})$ \cite{Laermann:2003cv}.\\
To study this aspect in the NJL model, we start from the parameters fit of \cite{Hatsuda:1994pi}, and we take the quark masses as free parameters; namely, we take the four and six fermion couplings fixed so as to reproduce the phenomenology of the realistical physical situation. This way of proceeding could seem as rather arbitrary, but for any value of the masses considered here, we have verified that the output parameters have reasonable values (critical temperatures $130\div200~\mbox{MeV}$, light quark scalar condensates $(-250\div-240~\mbox{MeV})^3$, constituent light quark masses $250\div350~ \mbox{MeV}$). Of course our results will be strongly model dependent, both for the choice of a specific set of parameters and of a particular model in itself. For instance, the NJL model provides an estimate of the position of the critical point at lower temperatures and higher chemical potentials with respect to those obtained in ladder-QCD and the ones from recent lattice simulation \cite{Fodor:2004nz}.\\
In Fig. \ref{fig:CondTms8} we show the behaviour of the light quark scalar condensates vs. temperature, in the $m_u=m_d=0$ limit, and for vanishing chemical potentials too. In the upper picture, $m_s$ is taken to be zero, and for a temperature of about $130~\mbox{MeV}$ there is a sharp first order transition. As we increase $m_s$ the discontinuity of the scalar condensates reduces (in the lower picture the case $m_s=8~\mbox{MeV}$ is shown), and when $m_s$ exceeds the critical value $m_s^C=10~\mbox{MeV}$, the zero temperature transition turns into a genuine second order transition. The value we get for $m_s^C$ is in any case smaller than the one from lattice predictions.\\
In Fig. \ref{fig:Diagmusmu} we plot the phase diagram in the ($\mu_q,m_s$) space, by taking $m_u=m_d$ fixed to zero, and $\mu_I=\mu_s=0$. The label $I/II$ indicates, for every couple $(\mu_q,m_s)$, whether, by increasing the temperature starting from zero, the transition is a first or a second order one. Actually, up to $m_s<m_s^C=10~\mbox{MeV}$, we have first order transition for every value of $\mu_q$, and consequently there is no critical point; for $m_s$ slight above the critical value, the critical point locates at a value of $\mu_q\sim200~\mbox{MeV}$. The critical value for $\mu_q$ grows together with $m_s$ until the strange quark decouples from the two light quarks and the critical point $\mu_q$ coordinate is independent on $m_s$, and lies at $\mu_q\sim~300\mbox{MeV}$.\\
Finally we have studied whether, in agreement with lattice analyses, a first order transition persists when we consider a non zero but small $m_u=m_d$: this does not happen in the NJL model with our choice of parameters, for any value of $m_s$. A recent work based on the linear sigma model had found the critical value for $m_u=m_d=m_s$ to be $m_{crit}=40\pm20\ \mbox{MeV}$ \cite{Herpay:2005yr}.


\subsection{Critical lines as a function of $N_f$}

Even though lattice analyses at finite density still present ambiguities in their different approaches, some general features about the critical line appear to be rather solid. 
In particular, if $T_0$ is the critical temperature for zero chemical potentials, the dependence of $(T/T_0)$ as a function of $(\mu/T_0)$ should be parabolic, (at least in the regime  $(\mu/T_0)<1$), i.e. of the form $(T/T_0)=1-\alpha (\mu/T_0)^2$.
Secondly, the $\alpha$ coefficient should depend on the number of flavors $N_f$, increasing with $N_f$; in fact, the curves relative to $N_f=2,2+1,3$ should be very close to each other and the one relative to $N_f=4$ should be steeper.\\
It is clear that a dependence of the $\alpha$ coefficient on $N_f$ must be related in any case with a coupling between the flavors; otherwise, the effective potential would become a sum of single flavors contributions, and the critical temperature would not depend on $N_f$. This fact can give us an idea about the strength of the coupling between the flavors at the phase transition, and of its possible reduction with temperature.
We will check the issue relative to the cases $2,2+1,3$ in the framework of the NJL model.\\
In this context, the aforementioned situations labeled by 2,3 are those with 2,3 massless flavors (in the case 2, $m_s$ is set to $5~\mbox{GeV}$ to decouple the strange quark); 2+1 is the physical case with realistic values of quark masses. In the following, $\mu$ will indicate a common value for the chemical potential equal for all the active flavors.

\begin{center}
\begin{figure}[htbp]
\includegraphics[width=14cm]{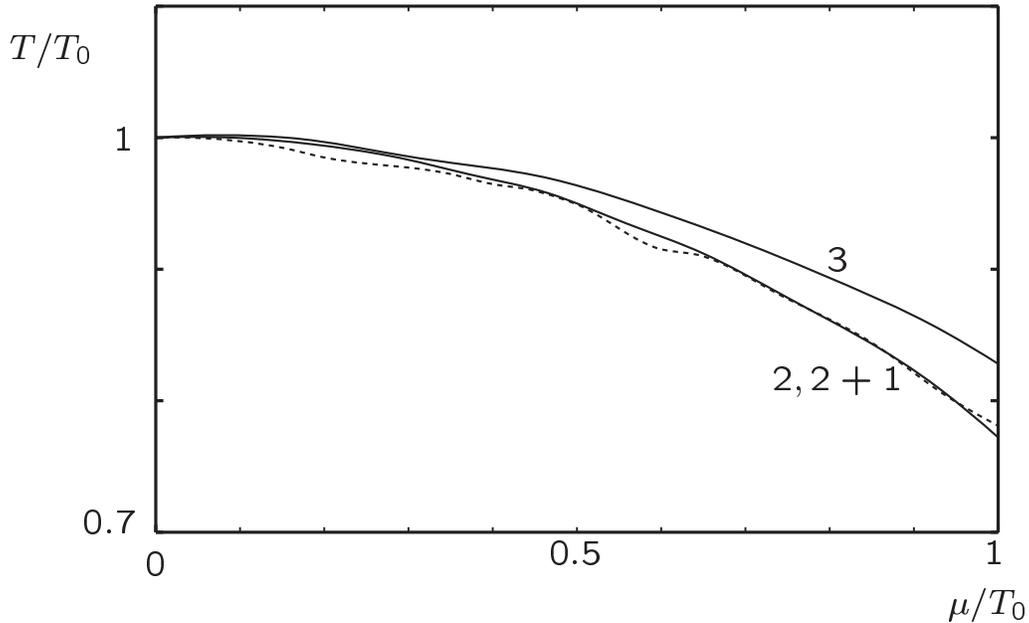}
\caption{\it Plot of the three cases $2,2+1,3$ in dimensionless units ($\mu/T_0,T/T_0$), for $\mu/T_0\leq1$. The curves relative to $2$ and $2+1$ (the dashed one) almost overlap. In disagreement with lattice simulations, the curve relative to $3$ stands slight above the others.} \label{fig:DiagTmuqadimflavors}
\end{figure}
\end{center}

In Fig. \ref{fig:DiagTmuqadimflavors} we show the phase diagrams relative to the cases $2+1,2,3$. Obviously, in case $2$, we are in the situation where $m_s<m_s^C$ and we have only first order transitions. As we are varying the quark masses to consider different situations, $T_0$ has a large range of variation (from $130$ to $200~\mbox{MeV}$); it is impressive that, as we plot the phase diagrams in dimensionless units, those large differences cancel out almost completely. This is the most striking evidence that our approach, in its simplicity, has some validity.

When showing the three cases on the same diagram, we can clearly observe the surprising overlap of the results.
The agreement between $2$ and $2+1$ cases (apart from numerical instabilities related with $2+1$ case) is remarkable, and good between $2$ and $3$ cases; the slope of case $3$, with respect to the lattice previsions, is smaller than that of $2$, but the difference is slight. \\

The results we get for $\alpha$ are the following: \\
\begin{equation}
~~2~~~~~~~~~~~\alpha=0.1995\pm0.025~~~~~~~~~~~~~~~~~~~~~~~~~~~~~~~~~~~~~~~~~~\nonumber
\end{equation}
\begin{equation}
~~2+1~~~~~~~\alpha=0.2496\pm0.0623~~~~~~~~~~~~~~~~~~~~~~~~~~~~~~~~~~~~~~~~~~
\end{equation}
\begin{equation}~~3~~~~~~~~~~~\alpha=0.1614\nonumber\pm0.011~~~~~~~~~~~~~~~~~~~~~~~~~~~~~~~~~~~~~~~~~~\end{equation}

The worse precision we obtain for the $2+1$ case depends on the larger error in the determination of the cross-over curve.\\

For completeness, we quote the results for $\alpha$ from lattice analyses: \\
\begin{eqnarray}&2&~~~~~~~\alpha=0.0507\pm0.0034;~\mu_u=\mu_d=\mu~ [6]\nonumber\\
&2&~~~~~~~\alpha=0.0504\pm0.0036;~\mu_u=\mu_d=\mu~ [8]\nonumber\\
&2&~~~~~~~\alpha=0.07\pm0.03;~\mu_u=\mu_d=\mu~ [10]\nonumber\\
&2+1&~~~~~~~\alpha=0.0288\pm0.0009;~\mu_u=\mu_d=\mu_s=\mu=\mu_B/3~ [4]\\
&3&~~~~~~~\alpha=0.0610\pm0.0009;~\mu_u=\mu_d=\mu_s=\mu=\mu_B/3~ [8]\nonumber\\
&3&~~~~~~~\alpha=0.114\pm0.046;~\mu_s=0~ [10]\nonumber\\
&4&~~~~~~~\alpha=0.099;~ \mu_f=\mu=\mu_B/3~ [6]\nonumber
\end{eqnarray}

\vskip0.5cm
We find that our predictions for $\alpha$ are bigger than those obtained by lattice approaches, apart from \cite{Ejiri:2003dc} for $N_f=3$; in that case, the results are comparable. A recent study based on a hadron resonance gas model \cite{Toublan:2004ks} give the result $\alpha=0.17\pm0.01$ and this value is in a good agreement with our results. A study within a chiral quark model give for $\alpha$ a value of about 0.1, extracted from Fig. 2 of \cite{Jakovac:2003ar}.

We have also studied the behaviour of the critical curve in ladder-QCD \cite{Barducci:1987gn} (in its version \cite{Barducci:2003un}); in this model there is no coupling between flavors, therefore it is independent on $N_f$.
We have found that the critical curve is flatter than that of NJL model, namely the coefficient $\alpha$ is much smaller, and hence closer to lattice predictions: $\alpha=0.0797\pm0.0056$.
We have attributed this feature to the lack of coupling between flavors in the model. On the other hand, since the value for the $\alpha$ coefficient we find in the NJL model is slightly higher than the value found in ref. \cite{Toublan:2004ks} and sensibly higher than the values obtained from other lattice analyses, we can argue that introducing an effective reduction of the $K$ coupling with temperature, as considered in ref. \cite{Hatsuda:1994pi,Costa:2005cz}, $\alpha$ would decrease at the meantime. Therefore, by considering the following temperature dependence of the 't Hooft term
\begin{equation}K(T)=K_0~\mbox{exp}(-T/T_1)^2\end{equation}  
we have verified that reducing $T_1$ the critical  temperature at $\mu=0$ is also reduced; in this way, the curve gets flatter. By taking two different values for $T_1$, and considering for simplicity the case $2$, we find the value $\alpha=0.1186\pm0.0061$ in the case $T_1=160~\mbox{MeV}$, and $\alpha=0.1084\pm0.0035$ in the case $T_1=100~\mbox{MeV}$.
In this way, the agreement with lattice is considerably improved; this can be considered an indirect proof of $U(1)_A$ effective restoration with temperature. 

For the sake of completeness, we have also studied the two flavor NJL model with the `t Hooft determinant. The model is not completely equivalent to the three flavor model in the limit of infinite $m_s$, since in the latter case a strange quark loop gives a contribution, proportional to the strange quark condensate, to the light quark constituent mass. In any case we do not expect a dramatic change in our results, with respect to previous expectations, for the behaviour of the critical curve.\\
We consider in this case the following expression for the interaction part of the Lagrangian (the $U(1)_A$ breaking determinant term can be rewritten as a four fermion interaction):

\begin{equation}
\mathcal{L}_{int}=G_1[(\bar{q}q)^2+(\bar{q}\vec{\tau}q)^2+(\bar{q}i\gamma_5q)^2+(\bar{q}i\gamma_5\vec{\tau}q)^2]+G_2[(\bar{q}q)^2-(\bar{q}\vec{\tau}q)^2-(\bar{q}i\gamma_5q)^2+(\bar{q}i\gamma_5\vec{\tau}q)^2]
\end{equation}
with
\begin{equation}
G_1=(1-\beta)G_0~~~;~~~G_2=\beta G_0
\end{equation}

The $\beta$ coefficient tells us how hard the flavor mixing is; it is maximal for $G_1=0$, namely for $\beta=1$.

For the choice of the parameters $G_1$ and $G_2$ we follow the approach proposed in ref. \cite{Hatsuda:1994pi,Frank:2003ve}.
In ref. \cite{Hatsuda:1994pi} the authors study the original two flavor Lagrangian, proposed by Nambu and Jona-Lasinio, with $G_1=G_2$ and therefore $\beta=0.5$. The value we find in this case for $\alpha$ is very similar with the result we obtained in the $SU(3)$ case in the limit $m_s\rightarrow\infty$: 
\begin{equation}\alpha=0.2107\pm0.0214~~~~~~~~~~~~~~~~~~~~~~~~~~~~~~\end{equation}

The authors of ref. \cite{Frank:2003ve} take $\beta$ as a free parameter instead. Here we furthermore consider the possible dependence of $G_2$ coefficient on the temperature, $G_2=G_2(T=0)~\mbox{exp} (-(T/T_1)^2)$.

If we take $G_2$ independent on the temperature (namely with $T_1=\infty$), the value we find for the $\alpha$ coefficient does not change by varying $\beta$: \begin{equation}\alpha=0.2142\pm0.0259~~~~~~~~~~~~~~~~~~~~~~~~~~~~\end{equation} again in agreement with previous analyses.
On the other hand, if we admit a restoration of $U(1)_A$ symmetry with temperature, we find a dependence on the $\beta$ coefficient.\\ For $\beta=0.2$ we have \begin{equation}\alpha=0.1484\pm0.009 \ \ \mbox{for} \ \ T_1=160~\mbox{MeV}~~~\end{equation} and \begin{equation}\alpha=0.1196\pm0.0161 \ \ \mbox{for}\ \  T_1=100~\mbox{MeV}.\end{equation}
For $\beta=0.3$ we have \begin{equation}\alpha=0.1024\pm0.0042\ \ \mbox{for} \ \ T_1=160~\mbox{MeV}\end{equation} and \begin{equation}\alpha=0.08101\pm0.007 \ \ \mbox{for} \ \ T_1=100~\mbox{MeV}.\end{equation}

However, according to Shuryak \cite{Shuryak:1993ee} it is very unlikely that restoration of $U(1)_A$ can occur before chiral symmetry restoration; therefore, the value $T_1=100\ \mbox{MeV}$ should not be taken too seriously. In any case, it appears clear that restoration of $U(1)_A$ symmetry can strongly influence the behaviour of the critical curve.

\label{sec:physics2}

\section{Conclusions}

In this paper we have studied some general features of the QCD critical line in the framework of a NJL model. In section one, we have compared the physics at $\mu_q\neq 0$ with that at $\mu_I\neq 0$. In section two, we have varied quark masses to show that the order of finite temperature transition changes if we consider small enough masses. In section three, we have studied the dependence of the slope of the critical curve $T_c$ vs. $\mu$ on $N_f$.
In recent times these questions have received much attention from the lattice community, due to the great improvement of finite chemical potential algorithms of simulation.

\begin{acknowledgments}
We wish to thank M.P. Lombardo and D. Toublan for useful discussions.
\end{acknowledgments}


\eject

\end{document}